# Model of Nanocrystal Formation in Solution by Burst Nucleation and Diffusional Growth

Daniel T. Robb and Vladimir Privman*

*Center for Advanced Materials Processing, Departments of Chemistry and Physics, Clarkson University, Potsdam, New York 13699-5721*

The phenomenon of burst nucleation in solution, in which a period of apparent chemical inactivity is followed by a sudden and explosive growth of nucleated particles from a solute species, has been given a widely accepted qualitative explanation by LaMer and co-workers. Here, we present a model with the assumptions of instantaneous rethermalization below the critical nucleus size and irreversible diffusive growth above the critical size, which for the first time formulates LaMer's explanation of burst nucleation in a manner allowing quantitative calculations. The behavior of the model at large times, $t$, is derived with the result that the average cluster size, as measured by the number of atoms, grows $\sim t$, while the width of the cluster distribution grows $\sim \sqrt{t}$. We develop an effective numerical scheme to integrate the equations of the model and compare the asymptotic expressions to results from numerical simulation. Finally, we discuss the physical effects which cause real nucleation processes in solution to deviate from the behavior of the model.

## 1. Introduction

Following Gibbs' foundational work,[1] classical nucleation theory was developed by Volmer and Weber, Becker and Döring, and other researchers to model the kinetics of subcritical clusters and the resulting nucleation rate. Volmer and Weber assumed[2] a thermal distribution of clusters below the critical size, whereas Becker and Döring formulated[3] the kinetics as a steady-state process, with monomer attachment and detachment resulting in a uniform rate of matter transport from single atoms up through the critical cluster size. In both approaches, the formation of the nucleated matter was assumed not to decrease locally the availability of monomers, and the nucleation rate remained steady.

During the burst nucleation of nanocrystals in solution, however, the explosive growth of nucleated particles implies that this assumption of constant atomic concentration is strongly violated. To understand the process, LaMer and Dinegar applied[4,5] classical nucleation theory to qualitatively describe the kinetics of burst nucleation. They proposed that, from a strong initial supersaturation, a rapid nucleation of particles would initially occur, followed by the absorption of diffusing atomic matter onto these nucleated particles. Their key observation was that the resulting reduction in supersaturation strongly depletes the local availability (concentration) of monomers, thus decreasing the rate of further nucleation, leading ultimately to a narrow size distribution of the nucleated, diffusively growing nanocrystals.

With continued improvement of instruments which probe the nanoscale, and the rapid growth of knowledge of nanobiological structures in recent years, the interest in industrial and biomedical applications of nanoparticles has increased markedly. Experimentalists continue to develop useful methods for nanoparticle preparation in a variety of systems,[6-10] adding to previously developed protocols.[11-14] Accurate, predictive quantitative modeling of burst nucleation, which could be of assistance in the further development of particle synthesis techniques, remains a theoretical challenge for several reasons. First, several kinetically coupled processes (monomer-producing chemical reactions, nucleation kinetics, and nanocrystal growth via attachment/detachment of monomers) are involved in burst nucleation experiments. In addition, Ostwald ripening may broaden the resulting distribution of nanocrystals, and further aggregation of these nanocrystals into secondary, polycrystalline colloid particles may occur. Considerable experimental[6,14-19] and modeling[6,18-24] effort has been reported on the latter, combined burst nucleation and further aggregation mechanisms that can yield uniform colloid particles. Second, experimental measurements of the kinetics of these processes are very difficult at the fast time scales and small particle sizes involved in burst nucleation and growth. Third, both thermodynamic and microscopic (kinetic) descriptions are needed to describe burst nucleation.

In this paper, we present a model of burst nucleation which assumes thermalization of clusters as long as they are below the critical size but diffusional growth of clusters large enough to have grown past the nucleation barrier. The model realizes the qualitative explanation by LaMer[4,5] for the first time in a formulation that allows actual calculations and is in this sense a quantitative model. We obtain analytical expressions for the large-time behavior of the model, which are compared to the results of numerical integration of the model equations, using representative physical parameters taken from recent work on gold nanoparticles.[6,23] The model should apply to the growth of any type of nanoparticle from a supersaturated solution, provided that the formation of critical nuclei can be described using a surface-plus-bulk free-energy expression (see section 2) and that further aggregation of already-nucleated particles (with each other) is not a significant effect. We discuss the limitations of the model assumptions and its relationship with other recent theoretical work.[25–30]

* To whom correspondence should be addressed.
  Web address: http://www.clarkson.edu/Privman
(1) Gibbs, J. W. *The Collected Works of J. Willard Gibbs*; Yale University Press: New Haven, CT, 1948.
(2) Volmer, M.; Weber, A. *Z. Phys. Chem.* **1926**, *119*, 227.
(3) Becker, R.; Döring, W. *Ann. Phys.* **1935**, *24*, 719.
(4) LaMer, V. K.; Dinegar, R. J. *J. Am. Chem. Soc.* **1950**, *72*, 4847.
(5) LaMer, V. K. *Ind. Eng. Chem.* **1952**, *44*, 1270.
(6) Privman, V.; Goia, D. V.; Park, J.; Matijević, E. *J. Colloid Interface Sci.* **1999**, *213*, 36.
(7) Peng, Z. A.; Peng, X. *J. Am. Chem. Soc.* **2001**, *123*, 183.
(8) Yu, W. W.; Peng, X. *Angew. Chem., Int. Ed.* **2002**, *41*, 2368.
(9) Teranishi, T.; Miyake, M. *Chem. Mater.* **1999**, *11*, 3414.
(10) Burda, C.; Chen, X.; Narayanan, R.; El-Sayed, M. A. *Chem. Rev.* **2005**, *105*, 1025.
(11) Turkevich, J.; Stevenson, P. C.; Hillier, J. *Discuss. Faraday Soc.* **1951**, *11*, 55.
(12) Petro, A. J. *J. Phys. Chem.* **1960**, *64*, 1508.
(13) Kerker, M.; Daby, E.; Cohen, G. L.; Kratohvil, J. P.; Matijević, E. *J. Phys. Chem.* **1963**, *67*, 2105.
(14) Matijević, E. *Chem. Mater.* **1993**, *5*, 412.
(15) Ocaña, M.; Morales, M. P.; Serna, C. J. *J. Colloid Interface Sci.* **1995**, *171*, 85.
(16) Ocaña, M.; Serna, C. J.; Matijević, E. *Colloid Polym. Sci.* **1995**, *273*, 681.
(17) Lee, S.-H.; Her, Y.-S.; Matijević, E. *J. Colloid Interface Sci.* **1997**, *186*, 193.
(18) Libert, S.; Gorshkov, V.; Privman, V.; Goia, D.; Matijević, E. *Adv. Colloid Interface Sci.* **2003**, *100–102*, 169.
(19) Libert, S.; Gorshkov, V.; Goia, D.; Matijević, E.; Privman, V. *Langmuir* **2003**, *19*, 10679.
(20) Dirksen, J. A.; Benjelloun, S.; Ring, T. A. *Colloid Polym. Sci.* **1990**, *268*, 864.
(21) Ring, T. A. *Powder Technol.* **1991**, *65*, 195.
(22) Dirksen, J. A.; Ring, T. A. *Chem. Eng. Sci.* **1991**, *46*, 2389.
(23) Park, J.; Privman, V.; Matijević, E. *J. Phys. Chem. B* **2001**, *105*, 11630.
(24) Privman, V. *Mater. Res. Soc. Symp. Proc.* **2002**, *703*, 577 (article T3.3).

Even with the simplifying assumptions, the equations of our model are nonlinear. We obtain however an analytical result at large times: We demonstrate that the distribution has an average cluster size (measured by the number of atoms) growing linearly with $t$ and a relative width shrinking as $t^{-1/2}$. In addition, because under the model assumptions the kinetics changes abruptly at the critical size, from the constraint of a thermal distribution below the critical size to irreversible diffusive growth above the critical size, numerical simulation of the model equations requires some care. We present a novel numerical method which deals effectively with the discontinuity in the kinetics.

This paper is organized as follows. In section 2, we present the governing equations of the model and discuss the approximations made in obtaining these equations. In section 3, an asymptotic solution of the equations is obtained analytically, from which the large-time behavior of the particle-size distribution is predicted. In section 4, a novel scheme for integrating the equations numerically is given, and numerical results are presented and compared with the asymptotic expressions from section 3. Finally, in section 5, we discuss the limits of applicability of the model for interpretation of experimental results, as well as the relationship of the model to recent theoretical work on the topic.

## 2. Model of Crystal Growth

Our model rests on many of the same assumptions made in a recent model[7,8,13,14,18] for *primary* particle production in two-stage colloid synthesis. We consider a supersaturated solution with monomer concentration $c$. Driven by thermal fluctuations, monomer aggregates (embryos) are produced, but their size is limited by the free-energy barrier imposed by the surface free energy, until one or more supercritical monomer aggregates (clusters, nanocrystals) are produced at a critical size, $N(c)$. For sizes above $N(c)$, the clusters are no longer in approximate thermal equilibrium, but they are assumed to grow irreversibly through the diffusion of monomers to their surfaces.

The true dynamics of the few-atom embryos involves the rather complicated transition rates between embryos of various sizes as well as possible internal restructuring processes, neither of which is well studied experimentally or theoretically. However, one often assumes that the dynamics of few-atom aggregates is very fast and leads to an approximately thermal distribution. This distribution can be modeled by the following form[1,6,23] of the free energy of an $n$-monomer embryo:

$$\Delta G(n,c) = -(n-1)kT \ln(c/c_0) + 4\pi a^2(n^{2/3}-1)\sigma \quad (1)$$

where $k$ is Boltzmann's constant, $T$ is the temperature in Kelvin, $c_0$ is the equilibrium concentration of monomers, and $\sigma$ is the effective surface tension. The first term is the free-energy contribution of the "bulk" of the embryo. Since it is negative for $c > c_0$, it favors larger clusters. The term is derived from the entropy of mixing of noninteracting monomers in solution, with the factor $\ln(c/c_0)$ ensuring that the bulk and solution phases are in equilibrium when $c = c_0$. The second, competing, positive term represents the surface free-energy cost, and it is proportional to the surface area of the embryo and therefore to $n^{2/3}$. The effective solute radius $a$, chosen so that the radius of an $n$-solute embryo is $an^{1/3}$, is defined by requiring that $4\pi a^3/3$ is the "unit cell" volume per monomer (including the surrounding void volume) in the bulk material.

We assume as usual that the distribution of embryo shapes can be neglected; that is, clusters are assumed spherical down to small $n$, yielding the form in eq 1. We note that even the surface tension of spherical particles is thought to vary with size, in a manner which is a topic of active research.[31] We neglect this effect as well as any geometrical factors that might be needed because real clusters are not precisely spherical. Note that the effective surface tension of nanoparticles is only partially understood at present, and the results of measurements have been found to vary somewhat with the measurement technique and chemical environment.[32] The surface tension is often taken[6,18,19] to be close to $\sigma_{bulk}$. Since $\sigma$ has been found[6,18,19,23] to have a strong effect on the resulting distribution, as well as the time scale of nucleation, it may be best however to fit it as a free parameter.

As the cluster size, $n$, increases, $\Delta G(n,c)$ increases up to the critical size,

$$N(c) = \left[\frac{8\pi a^2 \sigma}{3kT \ln(c/c_0)}\right]^3 = \left[\frac{2A}{3 \ln(c/c_0)}\right]^3 \quad (2)$$

where $A \equiv 4\pi a^2\sigma/kT$. Beyond the barrier, for $n > N$, the free energy decreases with $n$, but, as usual in nucleation theories, we assume that the kinetics becomes irreversible and is no longer controlled by $\Delta G$. The feature specific to burst nucleation is that the nucleation barrier depends on the monomer concentration, $c$, resulting in suppression of the nucleation after the initial burst, during which $c/c_0$ decreases by several orders of magnitude.

As indicated above, we assume that embryonic matter below the critical size $N$ is thermalized on a time scale much faster than that of the other dynamical processes in the system, so that the concentration of embryos of sizes in $(n, n + dn)$, given by $P(n,t) dn$, approximately follows a thermal distribution,

$$P(n,t) = c(t) \exp\left[\frac{-\Delta G(n,c(t))}{kT}\right] \quad (3)$$

where $c(t)$ is the time-dependent monomer concentration. The rate of nucleation is then approximated by[6]

$$\frac{dP(N+1,t)}{dt}\bigg|_{nucleation} = K_N c P(N,t) = K_N c^2 \exp\left[\frac{-\Delta G(N,c)}{kT}\right] \quad (4)$$

---

(25) Ludwig, F.-P.; Schmelzer, J. *J. Colloid Interface Sci.* **1996**, *181*, 503.
(26) Farjoun, Y.; Neu, J. C. 2007, arXiv:cond-mat/0702372v1. arXiv.org e-Print archive. http://arxiv.org/PS_cache/cond-mat/pdf/0702/0702372v1.pdf.
(27) Mozyrsky, D.; Privman, V. *J. Chem. Phys.* **1999**, *110*, 9254.
(28) Kelton, K. F.; Greer, A. L.; Thompson, C. V. *J. Chem. Phys.* **1983**, *79*, 6261.
(29) Kelton, K. F.; Greer, A. L. *Phys. Rev. B* **1988**, *38*, 10089.

(30) Shore, J. D.; Perchak, D.; Shnidman, Y. *J. Chem. Phys.* **2000**, *113*, 6276.
(31) Vanithakumari, S. C.; Nanda, K. K. *J. Phys. Chem. B* **2006**, *110*, 1033.
(32) Chernov, S. F.; Fedorov, Y. V.; Zakharov, V. N. *J. Phys. Chem. Solids* **1993**, *54*, 963.

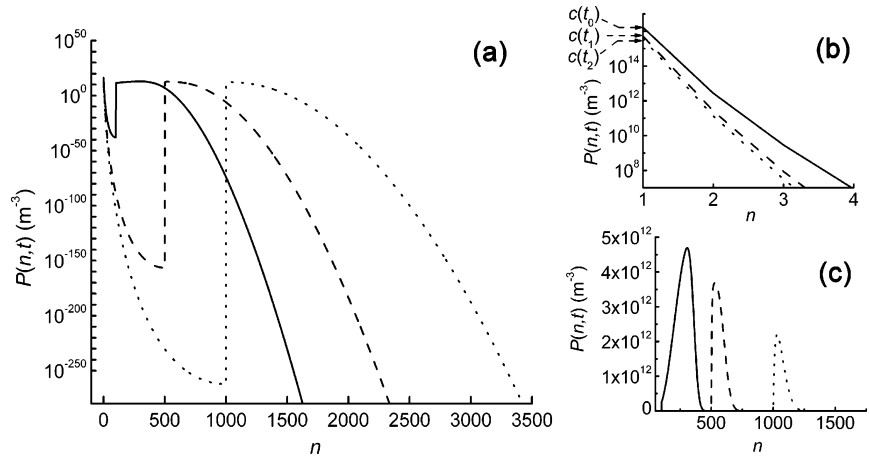

**Figure 1.** Cluster size distributions at times $t_0 < t_1 < t_2$, represented by solid, dashed, and dotted lines, respectively. (a) Semilog plot of a typical time evolution of the particle distribution, according to our model assumption of instantaneous rethermalization of clusters which are overtaken by the growing critical size, $N(c(t))$. The distributions change discontinuously, at the growing critical size, from a subcritical thermalized distribution to a peaked distribution of supercritical clusters. (b) Close-up at a very small cluster size, illustrating how the large decrease in the value $P(N(c(t)),t)$ of the distribution at the critical size, as seen in (a), is caused by a modest (5-fold) decrease in the monomer concentration, $c(t)$. (c) Plots (on the linear scale) of peaked distributions of supercritical clusters, which approach the form of a tail of a Gaussian as time increases.

where

$$K_n = 4\pi(a + an^{1/3})(D + Dn^{-1/3}) \approx 4\pi an^{1/3}D \quad (5)$$

is the Smoluchowski expression[33,34] for the rate of intake of diffusing monomers by spherical particles (we assume $n > N(c)$ $\gg 1$ for supercritical clusters) and $D$ is the diffusion coefficient for monomers in a solution with viscosity $\eta$; $D$ could be estimated as $\sim kT/6\pi\eta a$.[35] Note that $c \equiv c(t)$ and $N \equiv N(c) \equiv N(c(t))$ in eq 4.

We model the expected rapid growth of the supercritical ($n > N$) clusters within the approximation of *irreversible* capture of diffusing monomers (i.e., neglecting detachment) using the master equation,[6,23,24,33]

$$\frac{\partial P(n,t)}{\partial t} = (c(t) - c_0)(K_{n-1}P(n-1,t) - K_n P(n,t)) \quad (6)$$

The factor $(c(t) - c_0)$ is used in place of $c(t)$ so that the growth of clusters stops when the equilibrium concentration, $c_0$, is reached. In actuality, the variation of surface tension with particle radius mentioned above is accompanied by a variation of the effective equilibrium concentration with radius, which leads to Ostwald ripening.[36] This as well as other possible coarsening processes, such as cluster−cluster aggregation,[37] are neglected here because burst nucleation is expected[7] to be a much faster process. However, we note that such coarsening processes will gradually widen the particle distributions seen in experiment.

In addition to attachment and detachment of monomers, clusters can undergo the complex phenomenon of internal restructuring, the modeling of which for nanoscale clusters[38,39] is only in its early stages. Without such restructuring, the clusters would grow according to diffusion-limited aggregation or similar processes and could be fractals.[37,40,41] However, observations of the density and X-ray diffraction data of colloidal particles aggregated from burst-nucleated nanocrystalline subunits indicate that they have the polycrystalline structure and *density of the bulk*.[6,14] There is primarily experimental but also modeling evidence[6,23] that for larger clusters such restructuring leads to compact particles with smooth surfaces, which then grow largely irreversibly. In view of this, we assume that growing clusters restructure and become compact on time scales faster than those of growth by diffusional attachment.

The distribution $P(n,t)$ evolves as follows in our model. As pictured in Figure 1, at an early time $t_0$, for $n < N(c(t_0))$, $P(n,t)$ drops off sharply from its peak value of $P(1,t_0) = c(t_0)$ according to the thermalized distribution in eq 3. There is then a discontinuity at $n = N$, followed by a peak of supercritical clusters, which can be seen clearly in Figure 1c. As a function of time, we expect $N$ to grow, because monomers are consumed by supercritical clusters in the adsorption process represented by eq 6. As a result, as time progresses, the thermal distribution will decrease in value, as seen in Figure 1b, but will extend to larger values of $n$, "eroding" the leftmost part of the supercritical distribution in the process.

The rest of the supercritical distribution grows by the absorption of diffusing monomers, and, as argued in section 3, at large times will eventually assume the form of a tail of a narrow Gaussian for $n > N$. The distribution at later times in Figure 1c is relatively narrower and is approaching this Gaussian tail form, although it has not yet reached the asymptotic regime. Note that realistic distributions in experiment would have a more symmetrical peak, more evenly spread below and above the critical size, as discussed further in section 5. This is not the effect occurring in Figure 1c, in which peaks occur entirely above the critical size in a transient effect.

To obtain the time evolution of $c(t)$ and $P(n > N, t)$, we express the conservation of matter in the system as

$$\int_1^{N(t)} nc(t) \exp\left[\frac{-\Delta G(n,c(t))}{kT}\right] dn + \int_{N(t)}^\infty nP(n,t) \, dn =$$
$$\int_1^\infty nP(n,0) \, dn \quad (7)$$

(33) Smoluchowski, R. v. *Z. Phys. Chem.* **1917**, *29*, 129.
(34) Weiss, G. H. *J. Stat. Phys.* **1986**, *42*, 3.
(35) Dhont, J. K. G. *An Introduction to Dynamics of Colloids*; Elsevier Science: Amsterdam, 1996; p 81.
(36) Ostwald, W. *Z. Phys. Chem.* **1907**, *34*, 295.
(37) Family, F.; Landau, D. P. *Kinetics of Aggregation and Gelation*; North-Holland: Amsterdam, 1984.
(38) Lewis, L. J.; Jensen, P.; Barrat, J.-L. *Phys. Rev. B* **1997**, *56*, 2248.
(39) Baletto, F.; Ferrando, R. *Rev. Mod. Phys.* **2005**, *77*, 371 (section 5c).
(40) Witten, T. A.; Sander, L. M. *Phys. Rev. Lett.* **1981**, *47*, 1400.
(41) Witten, T. A.; Sander, L. M. *Phys Rev. B* **1983**, *27*, 5686.

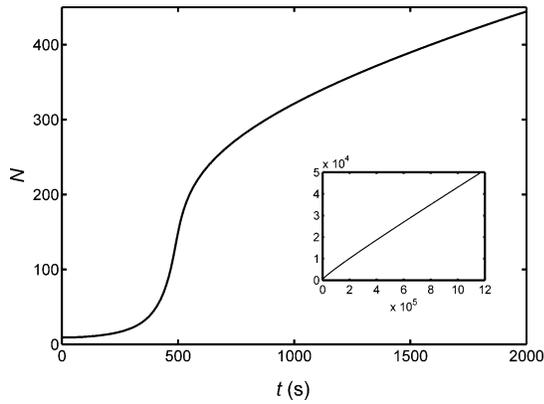

**Figure 2.** Plot of critical cluster size $N$ vs time $t$ during the initial nucleation burst for the model with parameter values given in section 3. The inset shows $N$ vs $t$ over the entire range of simulation times, illustrating the approximately linear behavior after the initial burst.

The large-time behavior is obtained (in section 3) by writing a continuous version of eq 6 and showing that it is solved approximately at large times by the (right-sided) tail of a moving Gaussian. In this asymptotic analysis, eq 7 is used only to determine the time-dependence of the "peak offset" between the peak of the moving Gaussian and the critical size, $N$. We then present (in section 4) a numerical integration scheme in which the discretization is appropriate to the discontinuous kinetics at the critical size. The more technical aspects of the derivations of sections 3 and 4 are given in the Supporting Information. We note that one must be consistent in the conventions for relating the discrete-$n$ quantities, such as the monomer concentration, which we took as $c(t) = P(1,t)$, to the values of the continuous distributions.

### 3. Large-Time Behavior of the Particle-Size Distribution

To motivate an ansatz for the large-time behavior of the model, we preview the results of numerical integration of the model equations, detailed in section 4. Most physical parameters in the simulation were taken equal to (or very close to) those used in previous simulations[6,23] of gold nanoparticles: $T = 293$ K, $a = 1.59 \times 10^{-10}$ m, $c_0 = 1.0 \times 10^{15}$ m$^{-3}$, and $D = 1.8 \times 10^{-9}$ m$^2$/s. However, we have used a lower supersaturation, because the model assumptions are likely better satisfied in such experimental systems (very large supersaturation leads to critical clusters of only a few atoms). In addition, the computational time required for integration of the model equations grows with increasing supersaturation. The surface tension was also chosen to be lower than that of gold, because the use of higher surface tensions necessitates smaller integration steps (increasing computational time requirements) and also results in the asymptotic behavior occurring at larger cluster sizes (increasing time and memory requirements). For the numerical results shown in this section, the initial monomer concentration and the surface tension were chosen as $c(t = 0) = 1.0 \times 10^{17}$ m$^{-3}$ and $\sigma = 0.183$ N/m, respectively. Past the critical nucleus size $N(t = 0) = N(c(t = 0)) = 9.0409$, found from eq 2, the initial cluster distribution was taken to be zero. (Results for seeded distributions are given in section 4.) In Figure 2, we plot $N(t)$ vs $t$. Following an induction time of approximately 400 s, there is a rapid increase in $N(t)$, which then crosses over slowly to the approximately linear dependence visible in the inset of Figure 2.

In Figure 3, we plot the cluster distributions $P(n,t)$ at times corresponding to the critical sizes $N(t) = 10\,000$, $20\,000$, and $50\,000$. The supercritical parts of the distributions are peaked at their respective critical sizes and fall off rapidly thereafter. As time progresses, the peak height of the distribution is seen to decrease, and the distribution width increases on an absolute scale but (as we will make explicit later) decreases relative to the critical size. Figure 4 illustrates that a part of a Gaussian curve, of the form

$$P_G(n,t) = \rho(t)c_0 \exp[-(\alpha(t))^2(n - K(t))^2] \quad (8)$$

with $\rho(t)$, $\alpha(t)$, and $K(t)$ as adjustable parameters, provides a good fit to the supercritical ($n > N(t)$) part of the distribution for $N = 20\,000$. We have found this to be true as well for fits to the rest of our numerical data for $N \geq 2000$. For $N \geq 5000$, the fitted distributions were observed to satisfy $K(t) < N(t)$, so that the numerical distributions are fit well at large times by the right-sided tail of a Gaussian. The tail is relatively narrow, that is, $1/\alpha(t) \approx 436 \ll K(t) \approx 19\,700$, for the fit shown in Figure 4 as seen in the inset.

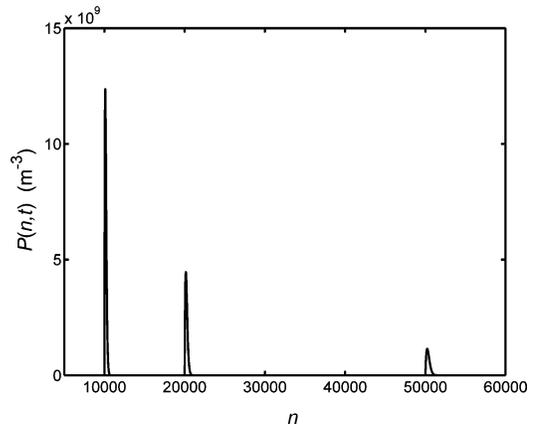

**Figure 3.** Supercritical probability distributions $P(n,t)$ vs cluster size $n$ for the same parameters as those in Figure 2. The thermalized parts of the distributions are not visible on the horizontal scale shown. Each peak constitutes a separate distribution, at times corresponding to the critical sizes $N = 10\,000$, $20\,000$, and $50\,000$. As $N$ increases, the distribution height decreases, the absolute width increases, while the relative width decreases.

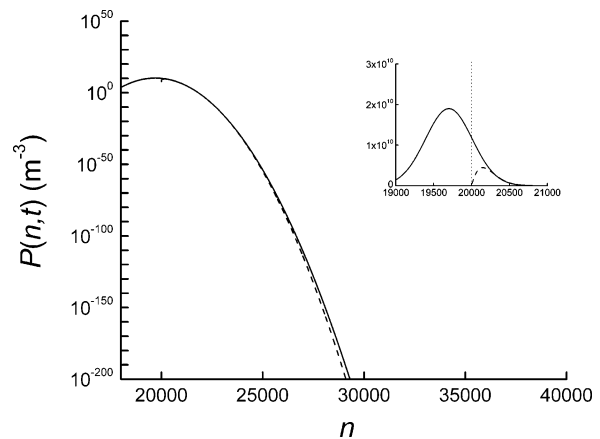

**Figure 4.** Plot of $P(n)$ vs $n$. The solid line shows the semilog plot of $P(n)$ vs $n$ at the time for which $N(t) = 20\,000$, while the dotted line shows the fitted Gaussian distribution $P_G(n) = 1.90 \times 10^{10} \exp[(-5.25 \times 10^{-6})(n - 19\,700)^2]$ in units of m$^{-3}$. (The relative standard errors estimated for the parameters $\rho(t)$, $\alpha(t)$, and $K(t)$ were approximately 10%, 0.5%, and 0.05%, respectively.) The inset shows a close-up of the fit on non-logarithmic scales.

We next employ the ansatz of a narrow Gaussian and characterize the asymptotic behavior of the variables $\alpha(t)$, $K(t)$, $\rho(t)$, and the "peak offset" $L(t) \equiv N(t) - K(t)$; these variables are illustrated graphically in Figure 5. First, we write the master equation, eq 6, in a continuous-$n$ form, keeping terms up to the second derivative,

$$K_{n-1}P_{n-1} \approx K_n P_n - \frac{\partial}{\partial n}(K_n P_n) + \frac{1}{2}\frac{\partial^2}{\partial n^2}(K_n P_n) \qquad (9)$$

where we have written $P_n$ for $P(n,t)$. Equation 6 becomes

$$\frac{\partial P_n}{\partial t} = (c - c_0)\left[\left(\frac{1}{2}\frac{\partial^2}{\partial n^2} - \frac{\partial}{\partial n}\right)(K_n P_n)\right] \qquad (10)$$

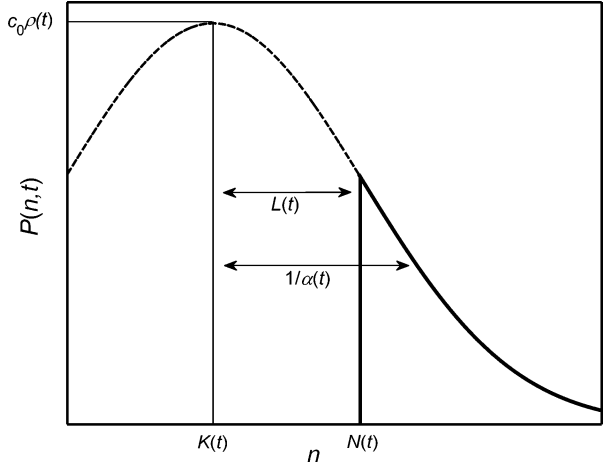

**Figure 5.** Schematic illustrating the parameters $N(t)$, $K(t)$, $L(t)$, $\rho(t)$, and $\alpha(t)$ associated with the Gaussian ansatz, eq 6, which is assumed in the asymptotic analysis of the model. The dashed line represents the entire fitted Gaussian, centered at $K(t) < N(t)$. The Gaussian tail at $n \geq N(t)$, which models the actual supercritical distribution, is represented by the thick solid line.

This equation describes the irreversible growth of clusters above the critical size $N(t)$, where, in keeping with the assumption of the narrow Gaussian, $P(n,t)$ takes on appreciable values only over a narrow range. Thus, we can approximate, for evaluation of the leading-order asymptotic behavior, $K_n \approx K_N = \kappa N^{1/3}/c_0$, where $\kappa \equiv 4\pi c_0 aD$.

Defining the dimensionless quantity $x(t) \equiv c(t)/c_0$, eq 10 becomes

$$\frac{\partial P_n}{\partial t} = \kappa(x(t) - 1)(N(t))^{1/3}\left(\frac{1}{2}\frac{\partial^2}{\partial n^2} - \frac{\partial}{\partial n}\right)P_n \qquad (11)$$

We can rewrite eq 2 as $x(t) = \exp[(2A/3)(N(t))^{-1/3}]$. Since, in the asymptotic (large-time) limit, $c(t) \to c_0$, that is, $x(t) \to 1$, we can approximate $x(t) - 1 \approx (2A/3)(N(t))^{-1/3}$. The factor $\sim N^{-1/3}$, which originates from the $n^{2/3}$ dependence of the surface energy in eq 1, then cancels the factor $\sim N^{1/3}$ in eq 11, which enters through the $n^{1/3}$ dependence of the diffusional growth rate in eq 5.

We define for later convenience the constant

$$z^2 \equiv \frac{4A\kappa}{3} = \frac{64\pi^2 a^3 \sigma c_0 D}{3kT} \qquad (12)$$

With this definition, eq 11 assumes a Fokker−Planck form in the particle-size space,

$$\frac{\partial P_n}{\partial t} = \frac{z^2}{2}\left(\frac{1}{2}\frac{\partial^2}{\partial n^2} - \frac{\partial}{\partial n}\right)P_n \qquad (13)$$

Substituting the Gaussian form, eq 8, into eq 13 produces three relations, which are found from equating the coefficients of the powers $n^2$, $n^1$, and $n^0$. From these relations, the large-time behaviors of the quantities $\alpha(t)$, $K(t)$, and $\rho(t)$ can be determined (see the Supporting Information, section S1):

$$\alpha \simeq 1/\sqrt{z^2 t}, \quad K \simeq z^2 t/2, \quad \rho \simeq \text{const}/\sqrt{z^2 t} \qquad (14)$$

Recall that the Gaussian distribution, with its peak at $n = K(t)$, is proposed as the asymptotic solution only for the cluster sizes $n > N(t)$. The asymptotic behavior of the peak offset $L(t) \equiv N(t) - K(t)$ is thus significant, and it can be shown (see Supporting Information, section S1) to be

$$L(t) \simeq \text{const}\sqrt{t \ln t} \qquad (15)$$

Therefore, the leading asymptotic behavior of the critical cluster size, $N$, is

$$N(t) \simeq z^2 t/2 \qquad (16)$$

The width of the distribution is given by $1/\alpha \simeq z\sqrt{t}$. We comment that the Gaussian distribution has provided a good quality fit for our numerical data at large times for various initial conditions, as detailed in the next section.

For completeness, we determine the asymptotic behavior of $c(t)$ from eq 16 and the relation

$$\frac{dN}{dt} = -\frac{3N}{c \ln(c/c_0)}\frac{dc}{dt} \qquad (17)$$

which follows directly from eq 2. Substituting $x = c/c_0 = 1 + \epsilon$ and $N(t) \simeq z^2 t/2$ into eq 17, we obtain

$$\frac{z^2}{2} = -\frac{3z^2 t c_0/2}{c_0(1 + \epsilon)\ln(1 + \epsilon)}\frac{d\epsilon}{dt} \longrightarrow -\frac{1}{3t} \approx \frac{1}{\epsilon}\frac{d\epsilon}{dt} \qquad (18)$$

which has the solution $\epsilon = \Lambda t^{-1/3} \Rightarrow c = c_0(1 + \Lambda t^{-1/3})$, with $\Lambda$ an integration constant.

## 4. Numerical Integration of the Model and Results for Seeded Distributions

In this section, we present and apply a novel and effective method for numerical integration of the model. Details are given in hopes that the approach will be of use in other situations with discontinuous kinetics. The assumption of discontinuous kinetics, which is an approximation of the real physical situation, creates technical difficulties in formulating the model entirely in terms of discrete particle sizes $n = 1, 2, ...$ Specifically, if written in discrete form, the conservation of matter, eq 7, could not be used to derive $dc/dt$ without significant ambiguities. The reason is that one would encounter derivatives of sums (over cluster sizes) with respect to a time-dependent summation index. The resulting time-dependence would not be continuous. For example, in a fully discrete formulation, the critical size would naturally be taken as $\lfloor N(t) \rfloor$, defined as the integer part of $N(t)$ as given in eq 2. When $c(t)$ decreased enough that the critical size $N(t)$ increased above the next whole integer, the embryonic matter at the former critical size, $\lfloor N(t - \Delta t) \rfloor$, would be absorbed (instantly) into the thermal distribution. With the addition of this matter, $c(t)$ could easily increase enough to return $\lfloor N(t) \rfloor$ to its previous value, $\lfloor N(t - \Delta t) \rfloor$ (or to an even lower value), in the process, creating an unphysical "matter gap," as illustrated in Figure 6. The influence of the unphysical matter gap would then propagate to higher cluster sizes.

To incorporate instantaneous rethermalization, irreversible diffusive growth, and a noninteger critical cluster size, we first make all the equations continuous and then rediscretize them in terms of the *distance above* the current critical cluster size $N(t)$. We use a data structure consisting of the monomer concentration $c(t)$, which also determines $N(t)$, and an array $S(m,t)$, $m = 1, 2, ...$, which stores the concentration of matter, $P(n,t)$, in the intervals $N(t) + m - 1 \leq n < N(t) + m$, $m = 1, 2, ...$, above the critical size. Since we know physically that $N(t)$ increases monotonically with time, we choose $N$ rather than the time $t$ as the variable of integration. Because we discretize the distance above the (time-varying) critical size, the equations turn out to be simpler with $N$ as the integration variable. In addition, there is some benefit gained in deciding on the integration step size, as discussed in the Supporting Information, section S2.

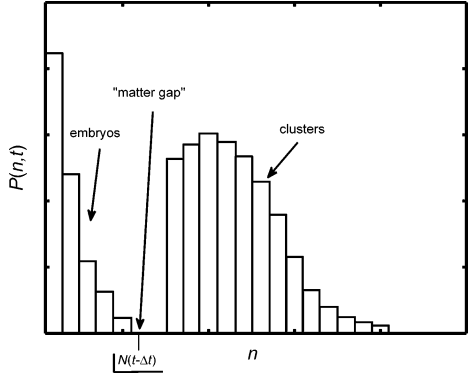

**Figure 6.** Illustration of the unphysical "matter gap" which could occur between the embryo and cluster distributions when the critical cluster size $N$ is treated as a discrete integer.

We now state the equations used to numerically simulate the model. The derivation of the equations can be found in the Supporting Information, section S3. To perform a simulation, we first specify the initial monomer concentration $c(t=0)$, which determines the initial value of $N$, as well as the initial cluster distribution $S(m,t=0)$ for $m = 1, ..., m_{max}$, where $m_{max}$ is an array size sufficiently large to hold the final cluster distribution which will be produced by the simulation. Note that $S(0,t)$ is defined at all times as the concentration of the largest embryo, that is, $S(0,t) \equiv c(t) \exp[-\Delta G(N(t),c(t))/kT]$.

Considering $c$ and $S(m)$ as functions of $N$, we increment $N$ by $\Delta N$ and want to update $c$ and $S(m)$ accordingly. From eq 2,

$$\Delta c = \frac{\partial c}{\partial N}\Delta N = -\frac{c \ln(c/c_0)}{3N}\Delta N \quad (19)$$

The derivatives of $S(m)$ are found to be

$$\frac{\partial S(m)}{\partial N} = [S(m+1) - S(m)] + \frac{(c - c_0)[K(m+N-1) S(m-1) - K(m+N) S(m)]}{\left(\frac{\partial N}{\partial c}\right)\left(\frac{dc}{dt}\right)} \quad (20)$$

where

$$\frac{dc}{dt} = -(c - c_0) \times \frac{\left(\sum_{m=1}^{m_{max}-1} K(m+N) S(m)\right) + (N+1) K(N) S(0)}{I_2(N) + N\frac{\partial N}{\partial c}[S(0) - S(1)]} \quad (21)$$

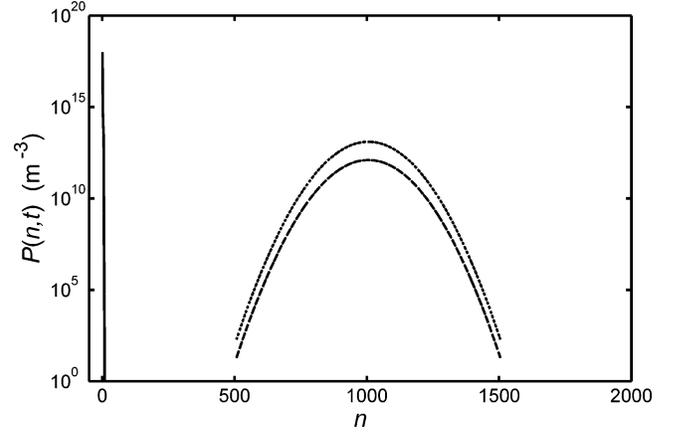

**Figure 7.** Semilogarithmic plot of the three initial conditions (ICs) A, B, and C used in the numerical integration in section 4. The thermal distribution is visible as the near-vertical (solid) line just above $n = 0$. IC A has no seeded supercritical clusters. ICs B (dashed line) and C (dotted line) have Gaussian distributions of seeded clusters centered at $\bar{n} = 1007$, with total masses $M_B = M_{th}$ and $M_C = 10 M_{th}$, respectively.

with $I_2(N) \equiv \int_1^N n^2 \exp[-\Delta G(n,c(N))/kT]\,dn$. Inserting eq 21 for $dc/dt$ into eq 20 gives

$$\frac{\partial S(m)}{\partial N} = [S(m+1) - S(m)] - \left(\frac{\partial c}{\partial N}I_2(N) + N(S(0) - S(1))\right) \times \frac{K(m+N-1) S(m-1) - K(m+N) S(m)}{\left(\sum_{m=1}^{m_{max}-1} K(m+N) S(m)\right) + (N+1) K(N) S(0)} \quad (22)$$

Finally, the time increment $\Delta t$ associated with the increment $\Delta N$ can be found from

$$\Delta t = \frac{\Delta c}{dc/dt} \quad (23)$$

with $\Delta c$ and $dc/dt$ given by eqs 19 and 21, respectively.

We examine the results obtained by numerical integration of these equations using three different initial conditions (IC), to be termed A, B, and C, with B and C corresponding to *seeded cluster distributions*. The parameters $c_0$, $T$, $a$, and $D$ were the same as those used in section 3. However, for all three ICs, the values $\sigma = 0.25$ N/m and $c(t = 0) = 1.0 \times 10^{18}$ m$^{-3}$ were used to illustrate the behavior at a somewhat higher supersaturation. The value of $z^2$, the constant introduced in eq 12, corresponding to these parameters is $z^2 = 0.0942$ s$^{-1}$. The initial distributions for the three ICs were (in addition to the "thermal" parts at $n < N$) as follows:

$$P_{A,B,C}(n > N(t=0), t=0) = \begin{cases} U_{A,B,C} \exp[-(n-1007)^2/100^2], & |n-1007| \leq 500 \\ 0, & |n-1007| > 500 \end{cases} \quad (24)$$

(The seeded distributions were chosen to be centered at $\bar{n} = 1007$ so that $\bar{n} - N(t=0) \approx 1000$.) We set $U_A = 0$, corresponding to no seeded distribution. The coefficients $U_B = 1.254 \times 10^{12}$ m$^{-3}$ and $U_C = 1.254 \times 10^{13}$ m$^{-3}$ were chosen so that the total masses, $M_{B,C} \equiv \int_{N(0)}^{\infty} n P_{B,C}(n, t=0)\,dn$, of the seeded distributions were $M_B = M_{th}$ and $M_C = 10 M_{th}$, respectively, where $M_{th} \equiv \int_1^{N(0)} n c(0) \exp[-\Delta G(n,c(0))/kT]\,dn$ is the mass of the initial thermal distribution. The three ICs are plotted in Figure 7.

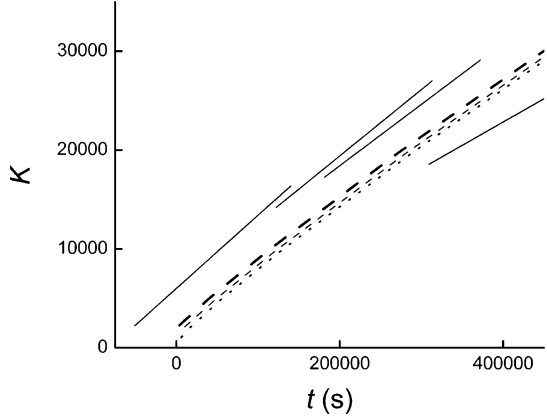

**Figure 8.** Plot of the peak location, $K$, of the fitted Gaussian distribution vs time, $t$, for numerical integration of the kinetic equations with parameters $\sigma = 0.25$ N/m and $c(t=0) = 1.0 \times 10^{18}$ m$^{-3}$. The dotted, thick dashed, and thin dashed lines correspond to the ICs A, B, and C, respectively. The upper solid lines show the slope $dK/dt$ for IC B at times corresponding to the critical cluster sizes $N = 5000$, $10\,000$, and $20\,000$ (the lines are centered at these times). The values of the slopes are given in Table 1. The asymptotic slope $z^2/2 = 0.0471$ s$^{-1}$ is represented by the lower solid line.

**Table 1. Values of the Local Slope $dK/dt$, in Units of s$^{-1}$, of the Peak Location of the Best-Fit Gaussian Distribution for ICs A, B, and C. The Values Are Shown for Times Corresponding to the Critical Sizes $N$**

| initial condition | $N = 5000$ | $N = 10\,000$ | $N = 20\,000$ | $N = 50\,000$ | $N = 100\,000$ |
|---|---|---|---|---|---|
| A | 0.0750 | 0.0669 | 0.0618 | 0.0578 | N/A |
| B | 0.0741 | 0.0671 | 0.0621 | 0.0578 | 0.0551 |
| C | 0.0739 | 0.0668 | 0.0619 | 0.0577 | 0.0549 |

Using eqs 19 and 21−23, we performed a numerical integration of the system with the three ICs A, B, and C. As in section 3, we then fit a Gaussian to the supercritical cluster size distributions, at times corresponding to the critical sizes $N = 2000$, $5000$, $10\,000$, $20\,000$, and $50\,000$, to determine the parameters $\rho(t)$, $\alpha(t)$, and $K(t)$. For ICs B and C, we also fit a Gaussian to the distribution at critical size $N = 100\,000$.

Figure 8 shows the time-dependence of the Gaussian peak location, $K(t)$, for all three ICs. The curve is well-approximated locally by a linear relationship, with a slope which decreases slowly with time. The slopes fitted at the times corresponding to $N = 5000$, $10\,000$, $20\,000$, $50\,000$, and $100\,000$ are shown in Table 1 for each of the ICs. It is clear from Table 1 that the IC has little effect on the asymptotic value of the slope and causes only an offset of the three $K$ vs $t$ curves, which can be discerned in Figure 8. The slopes are consistent with a slow convergence to the value $z^2/2 = 0.0471$ s$^{-1}$, as predicted by the asymptotic analysis in section 3. In Figure 9, we present a similar plot for the parameters used in section 3. The slopes $dK/dt$ (given in the caption to Figure 9) are consistent with a slow convergence to the predicted asymptotic slope for these parameters, $z^2/2 = 0.0345$ s$^{-1}$.

In Figures 10−12, we plot the curves $N_A(t)$, $N_B(t)$, and $N_C(t)$ from the three ICs A, B, and C, respectively, on progressively shorter time scales. In Figures 10 and 11, at $t \geq 2000$ s, the curves satisfy $N_B(t) − N_C(t) \approx N_C(t) − N_A(t) \approx 600$. The ordering $N_B(t) > N_C(t) > N_A(t)$ can be understood as follows. For IC A (with no seeded distribution), at the earliest times, $|dc/dt|$ is very small, since it is necessary to nucleate an appreciable cluster distribution before atomic matter can begin to aggregate significantly upon it. As the number of clusters begins to increase, the rate $|dc/dt|$ increases with it. The product $(c − c_0)\Sigma$, where $\Sigma$ is the total number (*not* the total mass) of supercritical clusters, is approximately proportional to the numerator of eq 21 and therefore also to the rate $dc/dt$. Since $c$ must decrease asymptotically to $c_0$ and the number of new clusters nucleated will drop sharply as $c$ approaches $c_0$, the product $(c − c_0)\Sigma$ and thus also $|dc/dt|$ must increase to a maximum and then decrease. This behavior is visible in the curve of $c$ vs $t$ for IC A plotted in Figure 13.

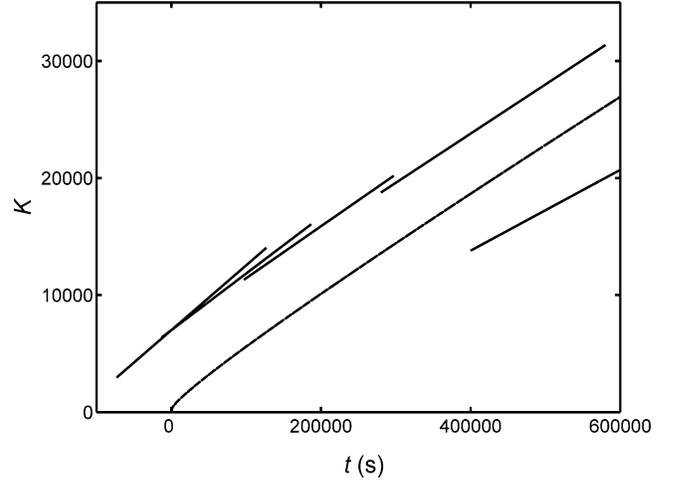

**Figure 9.** Plot of the peak location, $K$, of the fitted Gaussian distribution, vs time, $t$, for numerical integration of kinetic equations with parameters $\sigma = 0.183$ N/m and $c(t=0) = 1.0 \times 10^{17}$ m$^{-3}$ (dashed line). The slopes $dK/dt$ approximated at times corresponding to $N = 2000$, $5000$, $10\,000$, $20\,000$, and $50\,000$ were 0.0554, 0.0484, 0.0444, 0.0420, and 0.0400 s$^{-1}$, respectively. The first four slope values are represented by the upper solid lines. The lower solid line shows the predicted asymptotic slope $z^2/2 = 0.0345$ s$^{-1}$.

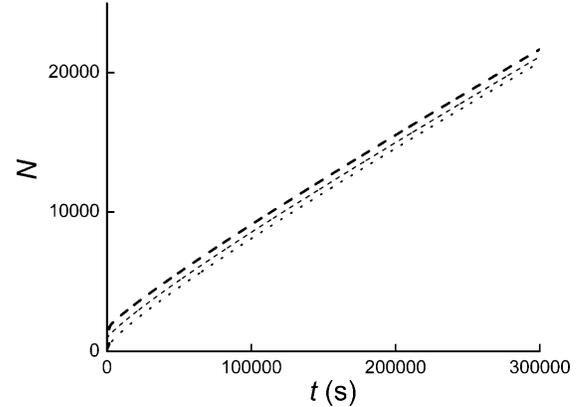

**Figure 10.** Plot of the critical size, $N$, vs time, $t$, on a long time scale for numerical integration of the kinetic equations with parameters $\sigma = 0.25$ N/m and $c(t=0) = 1.0 \times 10^{18}$ m$^{-3}$. The dotted, thick dashed, and thin dashed lines correspond to ICs A, B, and C, respectively. The data are shown for only one-sixth of the full integration time of $\sim 1.8 \times 10^6$ s, so that the lines can be distinguished.

The kinetics of the critical size $N$ can be understood in terms of the kinetics of $c$ through eq 17. While $|dc/dt|$ goes through its maximum, varying by less than an order of magnitude, the monomer concentration $c$ decreases by several orders of magnitude to relieve the supersaturation. Meanwhile, $N$ slowly increases, and $\ln(c/c_0)$ slowly decreases. Thus, at early times, the large factor $c$ in the denominator of the right-hand side of eq 17 overwhelms the effect of the other factors, which explains the extremely slow initial growth (induction) evident at early times in Figure 12. As $c$ decreases, $dN/dt$ increases significantly until $c$ reaches the vicinity of $c_0$, which for IC A occurs between $t =$

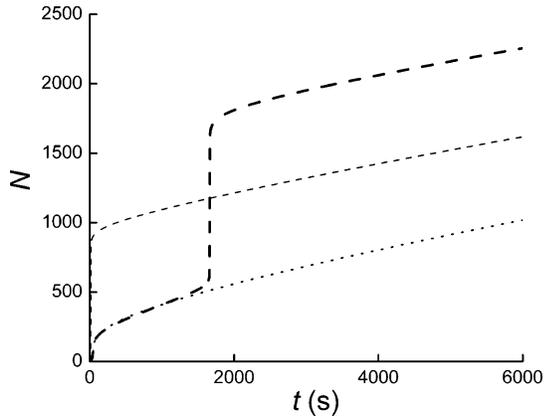

**Figure 11.** Plot of critical size, $N$, vs time, $t$, on an intermediate time scale for numerical integration of kinetic equations with parameters $\sigma = 0.25$ N/m and $c(t=0) = 1.0 \times 10^{18}$ m$^{-3}$. The dotted, thick dashed, and thin dashed lines correspond to ICs A, B, and C, respectively.

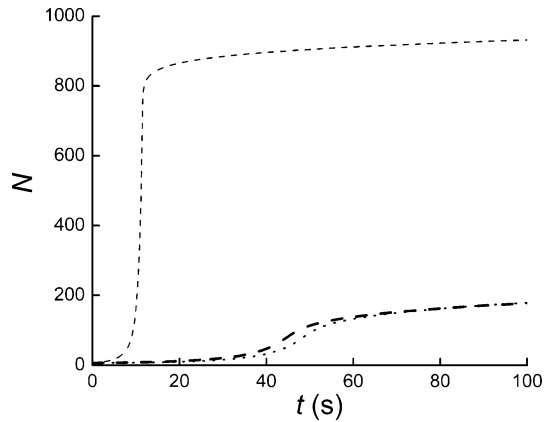

**Figure 12.** Plot of critical size, $N$, vs time, $t$, on a short time scale for numerical integration of kinetic equations with parameters $\sigma = 0.25$ N/m and $c(t=0) = 1.0 \times 10^{18}$ m$^{-3}$. The dotted, thick dashed, and thin dashed lines correspond to ICs A, B, and C, respectively.

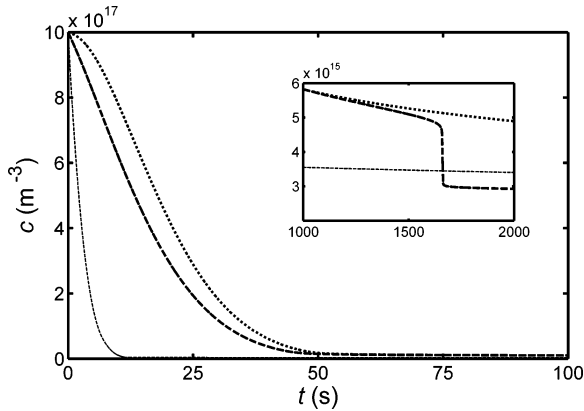

**Figure 13.** Plot of atomic matter concentration, $c$, vs time, $t$, on a short time scale for numerical integration of the kinetic equations with parameters $\sigma = 0.25$ N/m and $c(t=0) = 1.0 \times 10^{18}$ m$^{-3}$. The dotted, thick dashed, and thin dashed lines correspond to ICs A, B, and C, respectively. The inset shows $c$ vs $t$ over a later range of times.

50 s and $t = 60$ s (see Figures 12 and 13). There is then a crossover into the asymptotic regime, where $c - c_0 \ll c_0$.

The behavior of $c(t)$ and $N(t)$ for IC B can be understood in similar terms, but, as one might expect, the presence of the initial seeded distribution for IC B plays an important role in the early time kinetics. With the seeded distribution ($M \approx M_{th}$) present, the product $(c - c_0)\Sigma$ begins with an appreciable value. In addition, $\Sigma$ increases only slightly (relatively speaking) due to nucleation, while $(c - c_0)$ decreases, as in IC A. Thus, $|dc/dt|$ begins at (or very near) its peak value and decreases steadily, as seen in Figure 13, and the behavior of $N(t)$ up to $t \approx 1600$ s is similar to that of IC A.

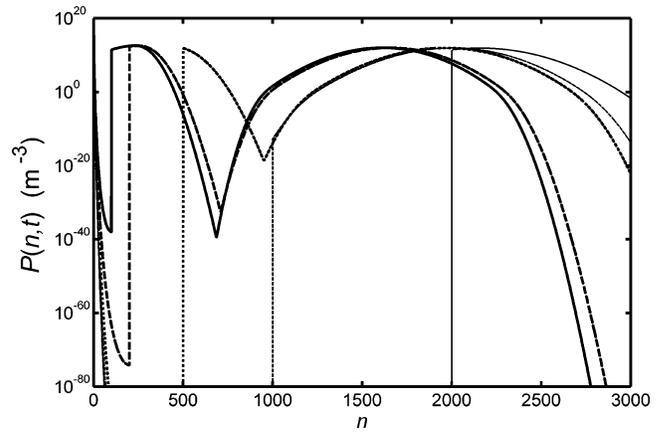

**Figure 14.** Plot of distribution $P(n,t)$ vs size $n$ for IC B, which specifies a seeded cluster distribution with $M \approx M_{th}$. The distributions are shown at times corresponding to the critical sizes $N = 100, 200, 500, 1000,$ and $2000$, which are represented by thick solid, thick dashed, dotted, thin dashed, and thin solid lines, respectively.

To understand the extremely rapid growth of $N$ between $t = 1600$ s and $1700$ s for IC B (cf. Figure 11), consider first the kinetics at large times. In that regime, the left edge of the supercritical cluster distribution (just above $N$) is consumed by the advancing thermalized distribution, while the remainder of the cluster distribution absorbs monomers. The analysis in section 3 established that when the supercritical cluster distribution has the shape of a tail of a Gaussian, these processes combine so as to produce a linear growth of $N$. We show the evolution of the cluster size distribution for IC B in Figure 14, in which the initial seeded distribution is clearly visible (compare to Figure 1). From Figure 14, it is clear that the cluster distribution at $N = 200$ ($t \approx 150$ s) is not shaped like a Gaussian tail. Because of the large seeded distribution, the difference between the amount of monomer matter absorbed and the amount of cluster matter consumed is larger than that in IC A, resulting in a slightly larger value of $|dc/dt|$ and a slightly larger value of $dN/dt$ (just visible in Figure 11 at $N = 200$). As the just-nucleated distribution is consumed, it is absorbed by the seeded distribution, which between $N = 200$ and $N = 500$ shifts strongly to higher $N$, as seen in Figure 14. When $N$ reaches the region of lower concentrations in the tail of the just-nucleated distribution, the difference between the amounts of absorbed and consumed matter increases dramatically, causing a rapid decrease in $c$ (and increase in $N$), until $N$ reaches a cluster size with enough matter present to restore the balance between absorbed and consumed matter. The assumption of the narrow Gaussian tail distribution then applies, and the kinetics approaches the asymptotic behavior.

The behavior for IC C can be understood in a similar fashion. In this case, as seen in Figure 13, the initial slope $|dc/dt|$ is much larger, since 10 times more matter exists in the seeded distribution to absorb the monomers. The decrease of $c$ and accompanying increase of $N$ are so rapid that no appreciable nucleated distribution can develop. This is apparent from the distributions at $N = 100$ and $N = 200$ in Figure 15, where the evolution of the cluster distribution for IC C is plotted. With the absorption of monomers far outpacing the consumption of cluster matter, the critical size $N$ again increases rapidly, until reaching a size ($N \approx 850$, see Figure 12) where the matter concentration is high enough to

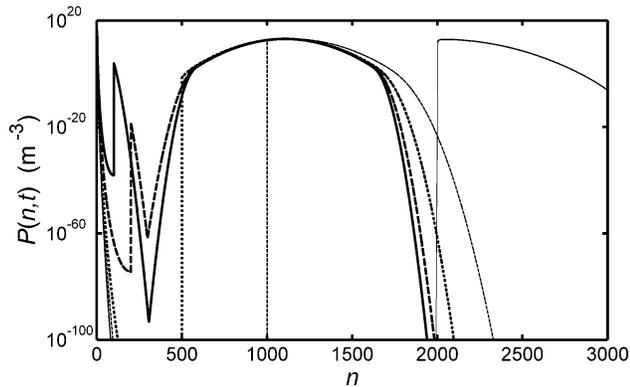

**Figure 15.** Plot of distribution $P(n,t)$ vs size $n$ for IC C, which specifies a seeded cluster distribution with $M \approx 10 M_{th}$. The distributions are shown at times corresponding to the critical sizes $N = 100, 200, 500, 1000$, and $2000$, which are represented by thick solid, thick dashed, dotted, thin dashed, and thin solid lines, respectively.

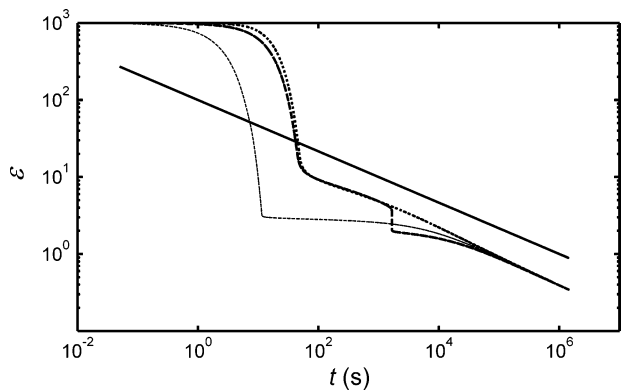

**Figure 16.** Log–log plot of $\epsilon(t) = c(t)/c_0 - 1$ vs $t$, illustrating the asymptotic relationship $\epsilon(t) \sim t^{-1/3}$. The dotted, thick dashed, and thin dashed lines correspond to the three ICs A, B, and C, respectively. The solid (reference) line is a plot of the function $100 t^{-1/3}$.

provide enough consumption to (nearly) balance the absorption and thus slow the decrease of $c$.

Finally, we present in Figure 16 a log–log plot of $\epsilon \equiv (c - c_0)/c_0$ vs $t$, which demonstrates that for all three initial conditions the asymptotic dependence is well fit by the $t^{-1/3}$ power law.

## 5. Discussion and Conclusion

From the asymptotic results for the critical size $N$ and the width $1/\alpha$ given in section 3, we can find the asymptotic large-time dependence of the average particle size and the relative width of the distribution in our model. Although the distribution consists of the right-sided tail of a Gaussian (cf. Figure 7), its width, $\Delta n$, is still given accurately by $\Delta n \sim 1/\alpha \sim \sqrt{t}$. Since $\Delta n/N \sim \sqrt{t}/t = t^{-1/2}$ decreases in time, the average particle size, $\bar{n}$, as measured by the number of monomers, is well approximated by the critical size, $\bar{n} \sim N \sim t$. The relative width thus has the dependence $\Delta n/\bar{n} \sim \sqrt{t}/t = t^{-1/2}$. Expressed in terms of the particle radius, $r$, the large-time dependences are $\bar{r} \sim t^{1/3}$ and $\Delta r/\bar{r} \sim t^{-1/2}$.

It is experimentally challenging to unambiguously quantify the size distribution of nucleated nanocrystals, because of their tendency to aggregate, their distribution of nonspherical shapes, the onset of Ostwald ripening, and other factors. Still, experimental distributions are commonly found to be two-sided (if not fully symmetric) around the peak, and the peak location is observed to stop increasing after a certain time.[7–9,11–13] For example, two-sided distributions have been measured in a gold hydrosol,[11] where the left side of the distribution was generally wider than the right, as well as in a sulfur hydrosol,[12] where the left side appeared clearly narrower. Recently, groups have achieved size control of very small CdO nanoparticles (3–6 nm diameter) using a noncoordinating solvent with varying concentrations of oleic acid,[7,8] and size control of Pt nanoparticles (2–4 nm diameter) using varying amounts of alcohol and protective PVP polymer in a water solution.[9] These protocols worked by altering the energetics of the nucleation process[7–9] and/or the kinetics of the diffusion and attachment process.[9] We note that for these extremely small particles the size distributions were indeed markedly right-sided,[7–9] as predicted by our model. In contrast, the earlier experiment on gold hydrosols[11] produced somewhat larger gold nanoparticles (10–25 nm diameter) via reduction of chloroauric acid with sodium citrate, and varied the size of the distribution simply through changes in reactant concentration and temperature. These parameters occur directly in our model. However, these experimental results[11] cannot be directly compared to our model in its current formulation, because, while we assume instantaneous availability of all monomers, the gold monomers are released slowly in the experiment. The slow release could be incorporated into our model, but it would require added assumptions about the rate of the reduction reaction.

The characteristic features of our model predictions (a one-sided distribution, and continued slow growth of the peak size (as $\bar{r} \sim t^{1/3}$)) can be attributed primarily to the assumption of instantaneous thermalization of clusters that fall below the critical size. At very small sizes, below a cutoff value which can be speculated to correspond to $n_{th} \approx 15-20$ building blocks[18,19,24,28–30] (here monomers), structures can evolve very rapidly, so that the assumption of fast, thermally driven restructuring is justified. At larger sizes, however, embryos can be expected to undergo a transition in which their internal atoms assume a more stable, bulklike crystal structure, and they no longer restructure as easily, except perhaps at their surface layers. Thus, at times for which $N(t) > n_{th}$, our model must be regarded as approximate, since it neglects the finite rate of dissolution/breakup of (larger) clusters below critical sizes $N > n_{th}$. It is clear that incorporating such a finite dissolution rate would add a left side to our distribution, bringing the model closer to agreement with aforementioned experimental observations for larger nanoparticles. In addition, with a two-sided peak and a finite dissolution rate of (larger) clusters below the critical size, the absorption of monomers should eventually be balanced by the detachment of monomers from the subcritical distribution (the left side), stopping the burst-nucleation stage and marking the beginning of the ripening stage.

Several groups have recently presented theoretical models attempting to describe the *combined* kinetics of nucleation and growth of nanocrystals in solution, using the ideas of classical nucleation theory. Ludwig and Schmelzer[25] inserted diffusion-limited rates for attachment and detachment into the expressions of Becker–Döring kinetics in an effort to numerically investigate von Weimarn's law in nucleation theory. They applied the resulting rate equations to the evolution above the critical size as well as to clusters below the critical size (where we note that the rate equations do not satisfy detailed balance). Farjoun and Neu[26] derived a novel expression for cluster growth rates by matching the diffusion flux to a kinetic expression of Becker–Döring form. Their eventual computational model applied to the evolution of a distribution of large clusters in the presence of small (but not too small) supersaturations, although the mechanism by which nucleated clusters grow to become large clusters seems somewhat unclear. In addition, Mozyrsky and Privman[27] formulated a radiation boundary condition to address the rates of the reactions occurring at the surface of clusters. Based on the success of a model of nucleation in a condensed system,[28,29] which assumes thermalization of the smallest clusters, as well as recent atomic-level simulations showing a disordered structure for the smallest of clusters (less than ~18 molecules) in solution,[30] we think that a promising strategy for future improvement of our model would be to utilize one of (or some combination of) the rate expressions developed in the above theoretical works.[25–27] It may also be worthwhile to attempt to incorporate the activation energy for the detachment of monomers from the surface of nanoparticles, as studied (for example) in a recent experiment on silica nanoparticles.[42]

In conclusion, we have presented a model for burst nucleation and growth of nanoparticles in solution, which realizes the qualitative explanation of LaMer[4,5] in a form which allows for analytical and numerical calculations. We have derived asymptotic predictions for the behavior of the average size and width of the particle distribution. In addition, we have presented a novel computational scheme for implementing the model and verified the predicted asymptotic behavior for three initial conditions, including cases with and without seed distributions. In assessing the model, we identified the assumption of instantaneous rethermalization as the main source of the expected discrepancy between the predictions of the model and experiments on nanocrystal growth in solution. Although the numerical results for the initial burst stage are likely to correspond well to experiment, the present model may be most interesting in the manner and extent to which it breaks down in comparison to experiment.

**Acknowledgment.** The authors gratefully acknowledge instructive discussions with D. V. Goia, E. Matijević, and I. Sevonkaev. This research was supported by the NSF under Grant DMR-0509104.

**Supporting Information Available:** Derivation of large-time behavior of the quantities $\alpha$, $K$, $\rho$, and $L$ in the Gaussian ansatz (section S1); discussion of integration times and choice of step-size for simulations (section S2); and derivation of equations used in numerical simulation (section S3). This material is appended.

(42) Rimer, J. D.; Trofymluk, O.; Navrotsky, A.; Lobo, R. F.; Vladhos, D. G. *Chem. Mater.* **2007**, *19*, 4189.

E-print arXiv:0707.2380 at http://www.arXiv.org

# Model of nanocrystal formation in solution by burst nucleation and diffusional growth

Daniel T. Robb and Vladimir Privman

**Supporting Information Section S1: Determining the large-time behavior of $\alpha, K, \rho,$ and $L$ in the Gaussian Ansatz.**

The asymptotic large-time behavior of the width $\alpha$, the peak location $K$, and the amplitude parameter $\rho$ of the Gaussian Ansatz (see Fig. 5) can be determined by inserting the assumed Gaussian form,

$$P_G(n,t) = \rho(t)c_0 \exp\left[-(\alpha(t))^2(n-K(t))^2\right], \quad (S1.1)$$

into the asymptotic form of the growth equation for the supercritical distribution,

$$\frac{\partial P_n}{\partial t} = \frac{z^2}{2}\left(\frac{1}{2}\frac{\partial^2}{\partial n^2} - \frac{\partial}{\partial n}\right)P_n . \quad (S1.2)$$

In the resulting expression, equating the coefficients of $n^2$ yields $-2\alpha^{-3}(d\alpha/dt) = z^2$, which has the solution $\alpha = (z^2 t + \Upsilon)^{-1/2}$, with $\Upsilon$ a constant. (The time dependence of $\alpha, K,$ and $\rho$ are suppressed for brevity). Equating coefficients of $n^1$ gives

$$4K\frac{d\alpha}{dt} + 2\alpha\frac{dK}{dt} = z^2\alpha - 2z^2 K\alpha^3 . \quad (S1.3)$$

Substituting $\alpha = (z^2 t + \Upsilon)^{-1/2}$ reduces Eq. (S1.3) to $dK/dt = z^2/2$. Thus the asymptotic form is the straightforward linear relationship $K = \Phi + \frac{z^2}{2}t$, with $\Phi$ a constant. Lastly, equating the $n^0$ (constant) coefficients, and simplifying, yields

$$\frac{d\rho}{dt} = \alpha\rho\left(2\frac{d\alpha}{dt}K^2 + 2\alpha K\frac{dK}{dt} + z^2\alpha^3 K^2 - \frac{z^2}{2}\alpha - z^2\alpha K\right) . \quad (S1.4)$$

Using $2d\alpha/dt = -z^2(z^2 t + \Upsilon)^{-3/2} = -z^2\alpha^3$ and $dK/dt = z^2/2$, Eq. (S1.4) simplifies to $d\rho/dt = -z^2\alpha^2\rho/2$, which has solution $\rho = \Omega(z^2 t + \Upsilon)^{-1/2}$.

The constants $\Upsilon, \Phi, \Omega$ cannot be determined by the asymptotic analysis. We summarize these large-time behaviors as

$$\alpha \simeq 1/\sqrt{z^2 t}, \quad K \simeq z^2 t/2, \quad \rho \simeq \Omega/\sqrt{z^2 t} . \quad (S1.5)$$

To determine the asymptotic behavior of the peak offset $L(t) \equiv N(t) - K(t)$, illustrated in Fig. 5, recall that total matter is conserved. Thus, as time increases, the amount of matter

$$M(t) \equiv \int_{N(t)}^{\infty} n P_G(n,t) dn \quad (S1.6)$$

in the supercritical cluster distribution must approach a constant value, equal to the initial total matter less the matter that remains in the thermal distribution as $c \to c_0$. Substituting Eq. (S1.1) for $P_G(n,t)$ into Eq. (S1.6) gives

$$M = \int_N^\infty n c_0 \rho \exp\left(-\alpha^2(n-K)^2\right) dn$$

$$= \frac{c_0\rho}{2\alpha^2} \int_{(\alpha L)^2}^\infty \exp(-u)\, du + \frac{c_0\rho K}{\alpha} \int_{\alpha L}^\infty \exp(-v^2)\, dv \ , \qquad (S1.7)$$

where the integral was rearranged by use of the change of variables $v = \alpha(n-K)$, and the definition $L = N - K$. (The time-dependences of $M$ and $L$ are suppressed for brevity.) It is straightforward to show that if the quantity $\alpha L$ does not diverge (as $t \to \infty$), then $M$ diverges; therefore, $\alpha L$ must diverge. To determine the nature of the divergence, we can replace the second integral in Eq. (S1.7) with the first term in the expansion, $\int_{\alpha L}^\infty \exp(-u^2)\, du \approx \exp(-\alpha^2 L^2)\left((2\alpha L)^{-1} - (4\alpha^3 L^3)^{-1} + \ldots\right)$. This results in the asymptotic expression

$$M = \frac{c_0\rho}{2\alpha^2} \exp(-\alpha^2 L^2)\left(1 + \frac{K}{L}\right). \qquad (S1.8)$$

Using the asymptotic forms, Eq. (S1.5), one can verify that $M$ approaches a constant if the peak offset at large times behaves as

$$L \simeq \Psi \sqrt{t \ln t}\left[1 - O\left(\frac{\ln\ln t}{\ln t}\right)\right]. \qquad (S1.9)$$

The constant $\Psi$ cannot be determined by the asymptotic analysis.

# Model of nanocrystal formation in solution by burst nucleation and diffusional growth

Daniel T. Robb and Vladimir Privman

**Supporting Information Section S2: Determining the integration times and choice of step-size for simulations.**

To determine a step size $\Delta N$ for the integration variable $N$ which prevents numerical instabilities, we can rewrite Eq. (22) in the approximate form

$$\Delta S(m) \approx -K_1 \cdot \Delta N \cdot \delta S(m) + \Delta N \cdot \delta S(m+1) \ . \qquad (S2.1)$$

In this equation, $\Delta S(m) \equiv S(m, N+\Delta N) - S(m, N)$, $\delta S(m) \equiv S(m, N) - S(m-1, N)$, and $K_1 = K(m+N)R$, where

$$R \equiv \frac{\frac{\partial c}{\partial N} I_2(N) + N(S(0) - S(1))}{\left(\sum_{m=1}^{m_{\max}-1} K(m+N)S(m)\right) + (N+1)K(N)S(0)} \ . \qquad (S2.2)$$

(The approximation that $-K(m+N)\delta S(m) \approx K(m+N-1)S(m-1) - K(m+N)S(m)$, used to write Eq. (S2.1), is quite mild since $K(n) \sim n^{1/3}$.)

Since Eq. (S2.1), like Eq. (6), is a discrete representation of the physical process of irreversible diffusional growth, any range of $m$ with oscillations, i.e., a range over which the quantity $\delta S(m)$ alternates in sign as $m$ increases, should be smoothed out by the time evolution. It can be shown that this will occur as long as $|K_1|\Delta N < 1$ and $\Delta N < 1$ in Eq. (S2.1). If either quantity is greater than one, then any oscillations present will grow quickly (and unphysically) in magnitude. To avoid this, we set the (variable) step size $\Delta N = [\Delta N]_{max} / |R| K(m_{max})$, with the additional restriction that $\Delta N < [\Delta N]_{max}$, using the value $[\Delta N]_{max} = 0.1$. With this choice, and with the array size taken as $m_{max} = 300000$, the numerical integration from $N = 0$ to 50000 took roughly 16 hours on our workstation, for each of the initial conditions. (Our computations were done on a Dell Precision 380 workstation with a 3.0GHz Pentium D processor.) A further integration from $N = 50000$ to 100000, performed for initial conditions B and C, took an additional 50 hours, the longer time due to the increasing size of the distribution being processed. The total mass (see Eq. (7)) was conserved to at worst 1% accuracy, which could be improved by choosing a smaller value for $[\Delta N]_{max}$, of course at the cost of longer computation time. The use of $N$ as the integration variable has the advantage that different initial supersaturations, which can lead to quite different time scales for the kinetics, can be handled with similar choices for $[\Delta N]_{max}$. However, because of the factor of $c$ present in the $\partial c / \partial N$ term in Eq. (S2.2), larger values of $R$ occur when integrating from higher initial supersaturations. This results in smaller integration steps $\Delta N$, and larger computation times.

# Model of nanocrystal formation in solution by burst nucleation and diffusional growth

**Daniel T. Robb**  and  **Vladimir Privman**

## Supporting Information Section S3: Derivation of equations used in numerical simulation.

We present here a detailed derivation of Eqs. (20) and (21), which are the results necessary to numerically integrate the model represented by Eqs. (6) and (7). The model incorporates instantaneous rethermalization below the critical size, irreversible diffusional growth above the critical size, and a non-integer critical cluster size. We will relate the monomer concentration $c(t)$, which also determines the non-integer critical size $N(t)$, to the quantities $S(m,t)$, $m = 1, 2, \ldots$, which represent the cluster concentrations in the size intervals $N(t) + m - 1 \leq n < N(t) - m$, $m = 1, 2, \ldots$. The critical size, $N$, will eventually be used instead of the time, $t$, as the integration variable. Until indicated otherwise in the following derivation, however, $c$ and $S(m)$ should be regarded as functions of the physical time, $t$. The time dependences of $c$, $N$, and $S(m)$ will be suppressed for brevity.

The conservation of matter, Eq. (7), written in terms of these quantities, reads

$$c I_1(t) + \Xi(t) = M \quad , \tag{S3.1}$$

where $I_1(t) \equiv \int_1^N n \exp[-\Delta G(n,c)/kT]\,dn$, and $\Xi(t) \equiv \sum_{m=1}^{\infty} S(m)[m+N]$. We differentiate the first term, $c I_1(t)$, obtaining

$$\frac{d}{dt}(c I_1(t)) = N c \exp\left[-\frac{\Delta G(N,c)}{kT}\right]\frac{\partial N}{\partial c}\frac{dc}{dt}$$

$$+ \int_1^N \left( n \frac{dc}{dt} \exp\left[-\frac{\Delta G(n,c)}{kT}\right] - \frac{nc}{kT}\frac{\partial[\Delta G]}{\partial c}\bigg|_{c(t)}\frac{dc}{dt}\exp\left[-\frac{\Delta G(n,c)}{kT}\right]\right)dn. \quad (S3.2)$$

From Eqs. (1) and (2), we can find directly

$$\frac{\partial N}{\partial c} = -\left[\frac{8\pi a^2 \sigma}{3kT}\right]^3 \frac{3}{c[\ln(c/c_0)]^4} = -\frac{3N}{c\ln(c/c_0)}, \quad (S3.3)$$

$$-\frac{1}{kT}\frac{\partial[\Delta G]}{\partial c}\bigg|_{c(t)} = \frac{n-1}{c}. \quad (S3.4)$$

Inserting these relations into Eq. (S3.2) and simplifying gives

$$\frac{d}{dt}(cI_1(t)) = \frac{dc}{dt}\left[Nc\frac{\partial N}{\partial c}\exp\left[\frac{-\Delta G(N,c)}{kT}\right] + I_2(t)\right], \quad (S3.5)$$

where $I_2(t) \equiv \int_1^N n^2 \exp[-\Delta G(n,c)/kT]\,dn$.

We next evaluate the time derivative of the sum $\Xi(t)$ in Eq. (S3.1), which is

$$\frac{d\Xi}{dt} = \sum_{m=1}^{m_{\max}}\left(\frac{\partial S(m)}{\partial t}[m+N] + S(m)\frac{\partial N}{\partial c}\frac{dc}{dt}\right), \quad (S3.6)$$

where a finite array size $m_{\max}$ has been substituted for the sum's upper limit. To proceed further, we use the following rate equation for $S(m)$, modified from Eq. (6) to include the movement of the boundaries of the cluster-size intervals represented by the quantities $S(m)$,

$$\frac{\partial S(m)}{\partial t} = (c-c_0)\left[K(m+N-1)S(m-1) - K(m+N)S(m)\right]$$

$$+ \frac{dN}{dt}\left[S(m+1) - S(m)\right]. \quad (S3.7)$$

The last term, containing $dN/dt$, describes the change in $S(m)$ resulting from the movement of the boundaries of the interval $N+m-1 \leq n < N+m$. At $m = m_{\max}$ in the sum Eq. (S3.6), we exclude the term $-(c-c_0)\left[K(m_{\max}+N)S(m_{\max})\right]$, in order not to lose matter from the system. Thus any matter which reaches size $m = m_{\max}$ accumulates there. For the purposes of this derivation, we set $S(m_{\max}+1)$, which appears in Eq. (S3.7) at $m = m_{\max}$, to zero; the term $S(m_{\max}+1)$ will not appear in the final equations. At $m=1$, we define $S(0) \equiv c\exp[-\Delta G(N,c)/kT]$.

Inserting Eq. (S3.7) into Eq. (S3.6) produces four terms, $E_1(t)$ through $E_4(t)$, which involve $(c-c_0)$ and $m$, $(c-c_0)$ and $N$, $dN/dt$ and $m$, and $dN/dt$ and $N$, respectively. Simplifying the telescoping sums which occur in these terms gives

$$E_1(t) = (c-c_0)\left(\sum_{m=0}^{m_{max}-1} K(m+N)S(m)\right), \qquad E_2(t) = (c-c_0)N \cdot K(N)S(0),$$

$$E_3(t) = \left[-\sum_{m=1}^{m_{max}} S(m)\right]\frac{dN}{dt}, \qquad E_4(t) = -\frac{dN}{dt}N \cdot S(1). \qquad (S3.8)$$

Summing these four terms, plus the last term in Eq. (S3.6), yields

$$\frac{d\Xi}{dt} = (c-c_0)\left(\left(\sum_{m=1}^{m_{max}-1} K(m+N)S(m)\right) + (N+1)K(N)S(0)\right) - \frac{dN}{dt}N \cdot S(1). \qquad (S3.9)$$

The full derivative of Eq. (S3.1) can then be written

$$\frac{d(cI_1)}{dt} + \frac{d\Xi}{dt} = \frac{dc}{dt}\left[N\frac{\partial N}{\partial c}(S(0)-S(1)) + I_2(t)\right]$$

$$+ (c-c_0)\left(\left(\sum_{m=1}^{m_{max}-1} K(m+N)S(m)\right) + (N+1)K(N)S(0)\right) = 0. \qquad (S3.10)$$

We then solve Eq. (S3.10) to find

$$\frac{dc}{dt} = \frac{-(c-c_0)\left(\left(\sum_{m=1}^{m_{max}-1} K(m+N)S(m)\right) + (N+1)K(N)S(0)\right)}{I_2(N) + N\frac{\partial N}{\partial c}[S(0)-S(1)]}, \qquad (S3.11)$$

which is identical to Eq. (21). We can obtain Eq. (20) by dividing Eq. (S3.7) by $dN/dt$.

# Model of nanocrystal formation in solution by burst nucleation and diffusional growth

Daniel T. Robb and Vladimir Privman

**TABLE OF CONTENTS GRAPHIC**

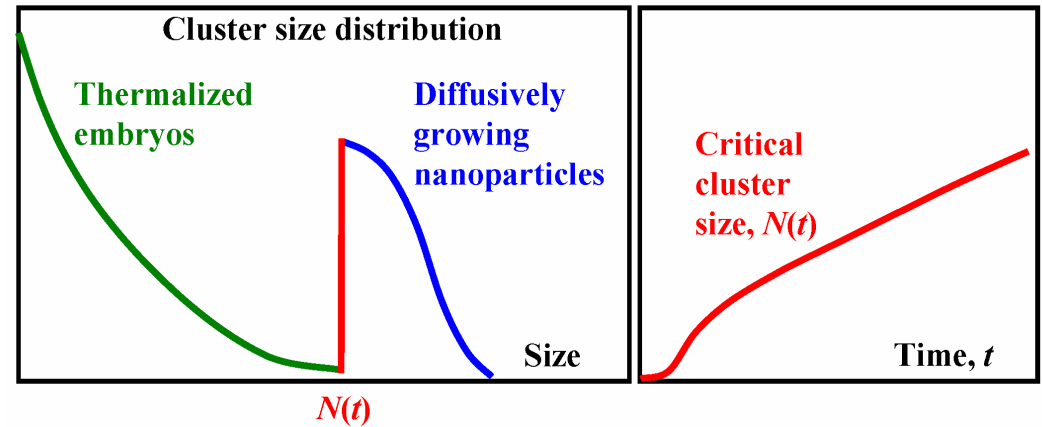